# OFDM based Sparse Time Dispersive Channel Estimation with Additional Spectral Knowledge

Hoomaan Hezaveh, Iman Valiulahi, and Mohammad Hossein Kahaei

*Abstract*-A new model for sparse time dispersive channels in pilot aided OFDM systems is developed by considering prior knowledge on channel time dispersions. Weighted atomic norm minimization is implemented in the model which enables a more accurate channel estimation. The channel response is identified by solving a Least Squares problem. In this work, we assume that time dispersions' associated frequencies can take any value with a minimum distance on the normalized interval [0,1). The performance of the new model is compared with conventional approaches. With respect to pilot number and SNR, the simulation results reveal that the new model performs superior to that of conventional methods. It is shown that both a lower energy and pilot number are required to achieve the same symbol error rate (SER) reported in previous works.

*Index terms*- Prior knowledge, Atomic norm minimization, sparse channel estimation, OFDM, pilot aided channel estimation.

## I. Introduction

Channel estimation is of great importance in wireless communication systems. Wireless channels suffer from the effect of time dispersion which causes a dramatic decrease in Signal to Noise Ratio (SNR) required for a fixed Symbol Error Rate (SER). The traditional approaches for channel estimation are based on Singular Value Decomposition (SVD). Despite the low complexity of these methods, information about the number of time dispersions is required in advance, which is not commonly known in practical scenarios [1-5]. On the other hand, Compressed Sensing (CS) has found its place as a sparse channel estimation technique in the recent years [6-10]. However, grid based dictionaries for compressive sampling suffer from basis mismatch [11-12]. Hence, Candes and Fernandez developed super resolution theory to resolve this problem [13-14]. [15] has investigated the proposed approach in [13] to compressed sensing regime. Furthermore, the performance of sparse time dispersive channel estimation problem in pilot aided OFDM systems [6] has been enhanced by adopting the approach of [15].

Due to slow changes in the channel in practical scenarios, time dispersions, which can be treated as delays, often occur in one or more fixed ranges [16-21]. This enables us to predict some ranges over which the channel time dispersions are located. In this work, in order to achieve a more exact estimation of the channel time dispersions, we limit our scope on these ranges and use them as prior knowledge. Here, prior knowledge is considered as a piecewise constant function which presents higher probability of occurrence in some ranges and lower probability of occurrence in others. This means that a new constraint can be imposed on the problem.

For a system with sufficiently slow changes in the channel response, the cumulative distribution function (CDF) can be presented as a probabilistic piece-wise constant function. This guarantees the existence of prior ranges and the associated probability of occurrence. For illustration, we consider the previously reported data for the case of three ranges of occurrence for the delay time [16] and apply a prior probabilistic piece-wise constant function to these data. The received signal level versus channel delay for different times is demonstrated in Fig.1. It can be inferred that the channel time dispersions are located in some predictable ranges. It is noteworthy that the arbitrary probabilistic piece-wise function specified to construct Fig.1, can be generalized to investigate more realistic scenarios.

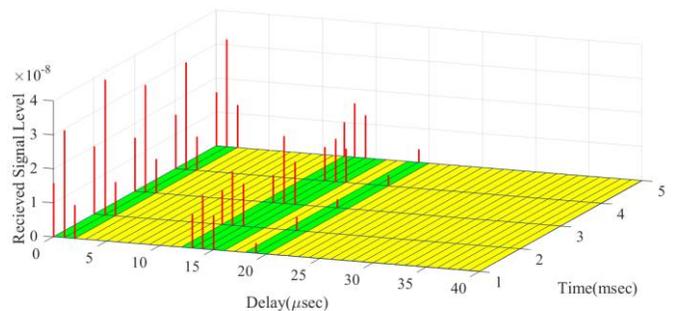

Fig. 1. Probabilistic prior ranges of occurrence for time dispersive channels. Green zones are associated with probabilistic prior ranges.

The existence of prior knowledge on channel time dispersions can be translated to statistical prior information on sparse spectral points in super resolution problem. Mishra et.al presented a new Semi-Definite Programming (SDP) to tackle this problem based on weighted atomic norm minimization [22]. We have implemented this approach to develop a new model for time dispersive channel estimation problem. The schematic of the model for estimating the channel response in systems with prior information on the channel parameters is shown in Fig.2.

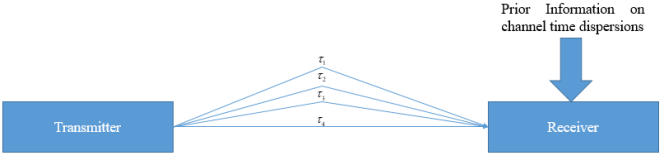

Fig. 2 New channel estimation model with prior information on time dispersions of the channel.

We impose a new constraint based on the ranges on atomic norm minimization problem. The proposed model is superior to conventional approaches in terms of SNR and pilot number needed for a fixed SER. Several simulation results are presented to support the main claim of the paper.

The rest of the paper is as follow: In Sec.II, the configuration of the OFDM system is presented followed by the atomic norm presentation and the associated SDP. Sec.III introduces weighted atomic norm and use it to present the new time dispersive channel response recovery problem. In Sec.IV simulation results are presented and discussed and Sec.V concludes the paper.

## II. PROBLEM FORMULATION

In OFDM systems N subcarriers are used to transmit data or pilots. After subjecting inverse discrete Fourier transform to data, cyclic prefix of length $L_{cp}$ symbols is added. After digital to analog conversion, OFDM continuous signal is sent through a wireless time dispersive channel. The baseband model of channel impulse response with $I$ scatters is [23]:

$$h_c(\tau) = \sum_{r=0}^{I-1} h_r \delta(\tau - \tau_r) \quad (1)$$

Where $h_r$ and $\tau_r$ are the complex gain and delay associated with $r$-th path and $\delta(\tau)$ represents Dirac delta function with one at $\tau = 0$. After analog to digital conversion and removing the cyclic prefix in receiver, the data is subjected to a multicarrier demodulator (discrete Fourier transform) to have:

$$y_0[n] = h[n]x[n] + w[n], n \in \mathcal{N} \quad (2)$$

Where $\mathcal{N} = \{k \mid 0 \leq k \leq N-1\}$, $x[n]$ is the sent data or pilots symbol at $n$-th subcarrier, $w[n]$ are i.i.d. zero-mean circularly symmetric complex Gaussian noise with variance $\sigma_n^2$ and $h[n]$ is discrete time Fourier transform of channel impulse response presented as:

$$h(n) = \sum_{r=0}^{I-1} h_r e^{-j2\pi \frac{n \tau_r}{N T_s}}, n \in \mathcal{N} \quad (3)$$

Where $T_s$ is the sampling interval and $j = \sqrt{-1}$. Consider $P$ pilots located at subcarrier positions $n_p, p = 0,...,P-1$. Thus by dividing received symbols $y[n_p]$ with equi-powered pilot symbols $x[n_p]$ we obtain:

$$y[n_p] = h[n_p] + \frac{w[n_p]}{x[n_p]} = \sum_{r=0}^{I-1} h_r e^{-j2\pi \frac{n_p \tau_r}{N T_s}} + w_1[n_p] \quad (4)$$

Where $w_1[n_p]$ are i.i.d. zero mean circularly symmetric complex white Gaussian noise with variance $\sigma_{np}^2$. According to the sparsity of the channel, we have $I \ll L_{cp}$ thus $P$ can be chosen such that $P < L_{cp}$. In order to reduce the number of pilots needed for correct reconstruction, we adopted the pilot allocation scheme of [6] and used $n'_p = \frac{L}{N} n_p, n_p \in \left\{0, \frac{N}{L},..., \frac{(L-1)N}{L}\right\}$ instead of $n_p$ where $L$ is chosen such that $L = \frac{N}{V}$ is an integer for integer $V$ ($\lceil L \rceil$ can be replaced if $L$ is not an integer. $\lceil L \rceil$ denotes the first integer greater than $L$). Therefore, $n'_p \in \mathcal{B} = \{0,...,L-1\}$ positions with $p = 0,...,P-1$ and $P < L$ pilots are used to estimate the channel response. Regarding above, channel estimation problem can be cast in the CS theory using the grid based algorithms. However, CS suffers from mismatching between true time dispersions and predefined grid. To overcome this issue, [15] suggested that one can estimate continuous sparse sources by atomic norm minimization. This approach was previously adopted in channel estimation problem [6]. The atomic norm is defined as:

$$\|\mathbf{x}\|_{\mathcal{A}} := \inf\{t > 0 : \mathbf{x} \in t\,conv(\mathcal{A})\}$$
$$= \inf_{\substack{c_r \geq 0 \\ \varphi_r \in [0, 2\pi) \\ f_r \in [0,1)}} \{\sum_r c_r : \mathbf{x} = \sum_r c_r \mathbf{a}(f_r, \varphi_r)\} \quad (5)$$

Where $t \in \mathbb{R}^+$, $conv(\mathcal{A})$ denotes convex hull of $\mathcal{A}$ and $\mathcal{A} = \{\mathbf{a}(f, \varphi) : f \in [0,1), \varphi \in [0, 2\pi)\}$ is the set of atom vectors $\mathbf{a}(f, \varphi)$ with components $a_n(f, \varphi) = e^{-j2\pi fn + \varphi}$, $n \in \mathcal{N}$. In off the grid CS, we only use part of $n$, e.g. $\mathcal{M} \subseteq \mathcal{N}$. According to the pilot allocation scheme used here, we can replace $\mathcal{N}$ with $\mathcal{B}$ and $\mathcal{M}$ with $\mathcal{R} \subseteq \mathcal{B}$. Thus, we would have the following atomic norm minimization problem to recover the signal of interest [15]:

$$\min_{\hat{\mathbf{x}}} \mu \|\hat{\mathbf{x}}\|_{\mathcal{A}} + \frac{1}{2} \|\mathbf{y} - (\mathbf{h})_{\mathcal{R}}\|_2^2 \quad (6)$$

where $\|.\|_2$ is $l_2$ norm, $\mu$ is a noise power dependent constant parameter, $\mathbf{y}$ is measurement vector and $(\mathbf{h})_{\mathcal{R}}$ is a vector which elements are defined according to set $\mathcal{R}$. One can solve (6) using the following SDP in [15] as below:

$$\begin{aligned}
&\underset{\mathbf{T}_L, \hat{\mathbf{x}}, t}{\text{minimize}} \quad \frac{1}{2|L|} tr(\mathbf{T}_L) + \frac{1}{2} t \\
&\text{subject to} \quad \begin{bmatrix} \mathbf{T}_L & \hat{\mathbf{x}} \\ \hat{\mathbf{x}}^* & t \end{bmatrix} \succeq 0, \\
&\qquad\qquad\quad \|\mathbf{y} - (\mathbf{h})_{\mathcal{R}}\|_2^2 \leq \mu,
\end{aligned} \quad (7)$$

where $tr(.)$ denotes trace of a matrix, $\mathbf{T}_N$ is a $N \times N$ positive semidefinite Toeplitz structured matrix, $\succeq 0$ denotes positive semi definiteness and $t \in \mathbb{R}^+$ is a constant. Using (7), one can find $\mathbf{T}_N$ and estimate $\tau_r$'s using a super resolution method like MUSIC or ESPIRIT. $h_i$'s are recovered by solving the following least squares problem:

$$\hat{\mathbf{h}}_I = \arg\min_{\mathbf{h}} \|\mathbf{Z}\mathbf{h}_I - \mathbf{y}\|_2^2, \quad (8)$$

Where $\mathbf{h}_I$ is an $I \times 1$ vector containing $h_r$'s and $\mathbf{Z}$ is a $P \times I$ matrix where $\mathbf{Z}[p,r] = e^{-j2\pi \hat{f}_r n_p} = e^{-j2\pi \frac{\hat{\tau}_r}{LT_s} n'_p}$, $r = \{1, ..., I\}$, $p = \{0, ..., P-1\}$.

## III. WEIGHTED ATOMIC NORM PRESENTATION

In this section, we present weighted atomic norm and use it to solve the new sparse time dispersive channel response recovery problem with prior information. There exist some prior information of channel time dispersions' probability distribution. In this work, we assume that our prior knowledge is in the form of a D-piece-wise constant function $D(\tau)$, where the value of this function in $[\tau_{L_i}, \tau_{H_i}], i = \{1, ..., D\}$, where $\tau_{L_i}$ and $\tau_{H_i}$ are the lower and higher bounds of $i$-th range, is $D_i$, and shows the existence probability of time dispersions. In order to incorporate the additional information to our problem, we use the same technique as [22] which is based on weighted atomic norm minimization. This norm is defined as below:

$$\|\mathbf{x}\|_{\mathcal{WA}} := \inf_{\substack{c_r \geq 0 \\ \phi_r \in [0, 2\pi] \\ f_r \in [0,1]}} \left\{ \sum_{r=0}^{I-1} \mathcal{W}_r |c_r| : x[n] = \sum_{r=0}^{I-1} c_r e^{-j2\pi f_r n}, n \in \mathcal{R} \right\} \quad (9)$$

Where $\mathcal{R}$ is the observation data set, and $\mathcal{W} = [\mathcal{W}_1, ..., \mathcal{W}_I]$ is a weight vector which entries are associated with the probability of occurrence of frequency $f_r$ which can be translated as delay $\tau_r$. According to the piece wise constant structure of $D(\tau)$, we have $\forall\{f_1, ..., f_r\} \in [f_{L_i}, f_{H_i}]$ or equivalently $\forall\{\tau_1, ..., \tau_r\} \in [\tau_{L_i}, \tau_{H_i}]$, $\mathcal{W}_1 = \cdots = \mathcal{W}_r = D_i$. Conversion between delays and normalized frequencies is pilot allocation scheme dependent. Here, pilot allocation scheme is as [6], therefore the conversion comes in the form of $f_r = \frac{\tau_r}{LT_S}$. We define the D-piece-wise constant weighted function $\Upsilon(\tau)$ so that $\Upsilon(\tau_i) = \frac{1}{D_i}$ where $i$ is corresponding to the interval $[\tau_{L_i}, \tau_{H_i}]$. To incorporate this novel structure into channel estimation, we suggest the following weighted convex optimization problem:

$$\begin{aligned}
&\min_{\hat{\mathbf{x}}} \quad \|\hat{\mathbf{x}}\|_{\mathcal{WA}} \\
&s.t. \quad \hat{x}[l] = x[l], \quad l \in \mathcal{R},
\end{aligned} \quad (10)$$

Where $x[l]$ denotes the $l$-th element of vector $\mathbf{x}$. Suggested problem is similar to the convex optimization problem in super resolution literature [22, III.2]. But SDP of (10) is not as (6). However, one can solve this problem by looking at its dual problem which can be obtained by Lagrangian theorem [24] as below:

$$\max_{\mathbf{q}} \langle \mathbf{q}_{\mathcal{R}}, \mathbf{x}_{\mathcal{R}} \rangle_{\Re}$$
$$s.t. \quad \|\mathbf{q}\|_{\mathcal{WA}}^{*} \leq 1, \quad (11)$$
$$\mathbf{q}_{\mathcal{B}\backslash\mathcal{R}} = 0,$$

Where $L \times 1$ vector $\mathbf{q}$ is dual variable, $\mathbf{q}_{\mathcal{B}\backslash\mathcal{R}}$ denotes relative complement of $\mathcal{R}$ with respect to set $\mathcal{B}$, $\langle .,. \rangle_{\Re}$ denotes the real part of the inner product and $\|\mathbf{q}\|_{\mathcal{WA}}^{*}$ is the dual norm of $\|\mathbf{q}\|_{\mathcal{WA}}$. Dual norm is defined as below:

$$\|\mathbf{q}\|_{\mathcal{A}}^{*} := \sup_{\|\hat{\mathbf{x}}\|_{\mathcal{A}} \leq 1} \langle \mathbf{q}, \hat{\mathbf{x}} \rangle_{\Re} = \sup_{f \in [0,1]} |\langle \mathbf{q}, \mathbf{a}(f,0) \rangle|. \quad (12)$$

According to the definition (12), the dual of weighted atomic norm is:

$$\|\mathbf{q}\|_{\mathcal{WA}}^{*} = \sup_{\|\hat{\mathbf{x}}\|_{\mathcal{WA}} \leq 1} \langle \mathbf{q}, \hat{\mathbf{x}} \rangle_{\Re}$$
$$= \sup_{\phi \in [0,2\pi], f \in [0,1]} \left\langle \mathbf{q}, \frac{1}{D(f)} e^{j\phi} \mathbf{a}(f,0) \right\rangle_{\Re} \quad (13)$$
$$= \sup_{f \in [0,1]} |\langle \mathbf{q}, \Upsilon(f) \mathbf{a}(f,0) \rangle|.$$

By substituting (13) in (11) we have:
$$\max_{\mathbf{q}} \langle \mathbf{q}_{\mathcal{R}}, \mathbf{x}_{\mathcal{R}} \rangle_{\Re}$$
$$s.t. \quad \sup_{f \in [0,1]} |\langle \mathbf{q}, \Upsilon(f) \mathbf{a}(f,0) \rangle| \leq 1, \quad (14)$$
$$\mathbf{q}_{\mathcal{B}\backslash\mathcal{R}} = 0.$$

According to the piecewise constant structure of $\Upsilon(f)$, we can expand (13) for different prior ranges:

$$\max_{\mathbf{q}} \langle \mathbf{q}_{\mathcal{R}}, \mathbf{x}_{\mathcal{R}} \rangle_{\Re}$$
$$s.t. \quad \sup_{f \in [f_{L_i}, f_{H_i}]} |\langle \mathbf{q}, \mathbf{a}(f,0) \rangle| \leq D_i, \quad (15)$$
$$\mathbf{q}_{\mathcal{B}\backslash\mathcal{R}} = 0,$$

For $i = \{1,...,D\}$. Note that $[f_{L_i}, f_{H_i}]$ and $\mathbf{a}(f,0)$ are the converted form of $[\tau_{L_i}, \tau_{H_i}]$ and $\mathbf{a}(\tau,0)$ respectively. We use the positive trigonometric polynomials technique suggested by [25] to convert the infinite dimensionality of dual problem into a linear matrix inequality problem as below:

$$\max_{\substack{\mathbf{q} \\ \mathbf{G}_{11}, \mathbf{G}_{12},...,\mathbf{G}_{1i} \\ \mathbf{G}_{21}, \mathbf{G}_{22},...,\mathbf{G}_{2i}}} \langle \mathbf{q}_{\mathcal{R}}, \mathbf{x}_{\mathcal{R}} \rangle_{\Re} - \sqrt{\sigma_{np}^{2}} \|\mathbf{q}_{\mathcal{R}}\|_{2}$$
$$s.t. \quad \mathbf{q}_{\mathcal{B}\backslash\mathcal{R}} = 0$$
$$\delta_{k_i} = \mathcal{L}_{k_i, f_{L_i}, f_{H_i}}(\mathbf{G}_{1i}, \mathbf{G}_{2i}),$$
$$k_i = 0,...,(L-1), \quad (16)$$
$$\begin{bmatrix} \mathbf{G}_{1i} & \frac{1}{\mathcal{W}_i}\mathbf{q} \\ \frac{1}{\mathcal{W}_i}\mathbf{q}^{*} & 1 \end{bmatrix} \succeq 0,$$

Where,
$$\mathbf{G}_{11}, \mathbf{G}_{12},...,\mathbf{G}_{1i} \in \mathbb{C}^{L \times L},$$
$$\mathbf{G}_{11}, \mathbf{G}_{12},...,\mathbf{G}_{1i} \in \mathbb{C}^{(L-1) \times (L-1)},$$

for $i = \{1,...,D\}$, $\delta_0 = 1$ and $\delta_k = 0$ for $k \neq 0$,

$$\mathcal{L}_{k, f_L, f_H}(\mathbf{G}_1, \mathbf{G}_2) = tr(\Theta_k \mathbf{G}_1)$$
$$+ tr((d_1 \Theta_{k-1} + d_0 \Theta_k + d_1^{*} \Theta_{k+1}).\mathbf{G}_2) \quad (17)$$

in which $\Theta_k$ is an elementary Toeplitz matrix with ones on its $k$-th diagonal and zeros elsewhere. Here, $k = 0$ corresponds to the main diagonal, and takes positive and negative values for upper and lower diagonals, respectively. $d_0$ and $d_1$ are defined as below:

$$d_0 = -\frac{\alpha\beta + 1}{2}, \quad (18)$$

$$d_1 = \frac{1 - \alpha\beta}{4} + j\frac{\alpha + \beta}{4}, \quad (19)$$

and

$$\alpha = \tan\frac{\omega_L}{2}, \quad (20)$$

$$\beta = \tan\frac{\omega_H}{2}, \quad (21)$$

Where $[\omega_L, \omega_H] \subset [-\pi, \pi]$ are defined as follow:

$$\omega_L = \begin{cases} 2\pi f_L & : 0 \leq f_L \leq 0.5 \\ 2\pi(f_L - 1) & : 0.5 \leq f_L \leq 1 \end{cases}, \quad (22)$$

$$\omega_H = \begin{cases} 2\pi f_H & : 0 \leq f_H \leq 0.5 \\ 2\pi(f_H - 1) & : 0.5 \leq f_H \leq 1 \end{cases}. \quad (23)$$

By solving (16), $F = [f_1, ..., f_I]$ frequencies or equivalently $T = [\tau_1, ..., \tau_I]$ delays can be found. Then, $h_i$'s are recovered by solving the same structured problem as (8). Finally, estimated channel response can be obtained by substituting $\hat{h}_r$'s and $\hat{\tau}_r$'s in (24):

$$\hat{h}[n] = \sum_{r=0}^{I-1} \hat{h}_r e^{-j2\pi \frac{\hat{\tau}_r}{NT_s} n_p} \quad (24)$$

Algorithm 1 summarizes the above discussion.

---

**Algorithm 1 :** Time dispersive sparse channel response recovery with probabilistic priors

---

1: Solve (15) and find $\mathbf{q}$.

2: Unknown frequencies can be identified by the magnitude of dual polynomial which achieves $D_i$ in $i$-th interval, mathematically, $\left| \langle \mathbf{q}^*, e^{j2\pi f_r n'_p} \rangle \right| = D_i$

$\forall f_r \in [f_{L_i}, f_{H_i}] \subseteq [0,1], i = \{1, ..., D\}, r = \{0, ..., I-1\}$.

3: Consider $f_r$ as $\frac{\tau_r}{LT_s}$ and find $\tau_r$'s.

4: Solve (8) and find $\hat{\mathbf{h}}_I$.

5: Substitute $\hat{h}_r$'s and $\hat{\tau}_r$'s in (24) to find estimated channel response.

---

Note that above formulation was made on the hypothesis of $D(\tau)$ being interval constant coefficients. Formulating another SDP for continuous weighting strategy is left as future work. To the best of author's knowledge this is the first work considering prior knowledge in OFDM based sparse time dispersive channel estimation problem in the continuous domain and presenting a new model as Fig.2 to improve channel response recovery. Several numerical simulations were made to investigate different aspects of the problem.

## IV. NUMERICAL RESULTS

To evaluate the performance of the proposed model we carried out MATLAB simulations. We compared the Mean Square Error (MSE) of this new model with the MSE of the one used in [6]. The difference is that here we use (10) as our objective function instead of (6). This origins from different model and assumptions we use here. The simulations are divided to three groups: first group focus on the number of pilots' effect on reconstruction MSE. According to the pilot allocation scheme used here, we compared this model with the one used in [6] known as AtomSR in terms of pilot number. Second group is devoted to SNR investigation. It is of great importance to see if the proposed model can help saving energy while SER remains the same. Subspace decomposition method, such as MUSIC, as a high resolution spectral estimation method is compared in terms of MSE for different SNRs. Third group focus on minimum frequency separation which, in terms of channel estimation, can be translated to minimum delay separation that this new model can distinguish. To carry out the simulations, we used CVX [26].

### A. Pilot Number

We considered two different ranges for delays to be located in, [9.6, 19.2] $\mu$sec & [44.8, 54.4] $\mu$sec. Delays can be translated to frequencies as [0.15, 0.30] & [0.70, 0.85] (Frequencies are normalized to range [0 1)). We considered a 2-piece-wise constant function as our prior probabilistic distribution function which is 1 at mentioned ranges and zero in other ranges. In this case $D_1 = 1$ & $D_2 = 1$ which means that our prior experiments identified delays only in these two bands. OFDM system assumed to have $N = 512$ subcarriers and maximum pilot number of $L = 64$. Pilot allocation scheme is as the one used in [6]. SNR is fixed at 20dB. In each simulation run different channel response with uniform distributed random frequencies in prior ranges and random complex coefficients with same prior and $I = 4$ paths was realized. This process was iterated for 1000 times. Note that at least $3I$ samples has to be taken and minimum frequency separation should be $\frac{4}{N}$ [15]. We used (25) as our comparison criterion:

$$MSE = \mathbb{E}\{\frac{1}{N} \sum_{n=0}^{N-1} \left| \hat{h}[n] - h[n] \right|^2\}. \quad (25)$$

Fig.3 shows the results for pilot number variation from 24 to 42. The results showed that using prior assumptions can greatly improve the performance of channel estimation with different pilot numbers. For better illustration, Fig.4-(a) depicts the dual polynomial and true frequencies in one iteration. Probabilistic prior ranges are shown with the yellow zone.

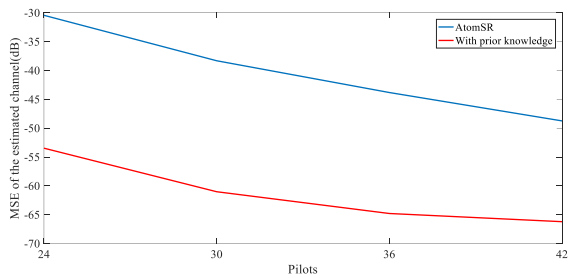

Fig. 3. MSE of the estimated channel with AtomSR method assumptions compared with presented model in the terms of pilot number. Pilot allocation scheme is the same as the one used in [6] and SNR is fixed at 20 dB. It can be deduced that the new model with prior assumptions greatly improves the performance of channel estimation.

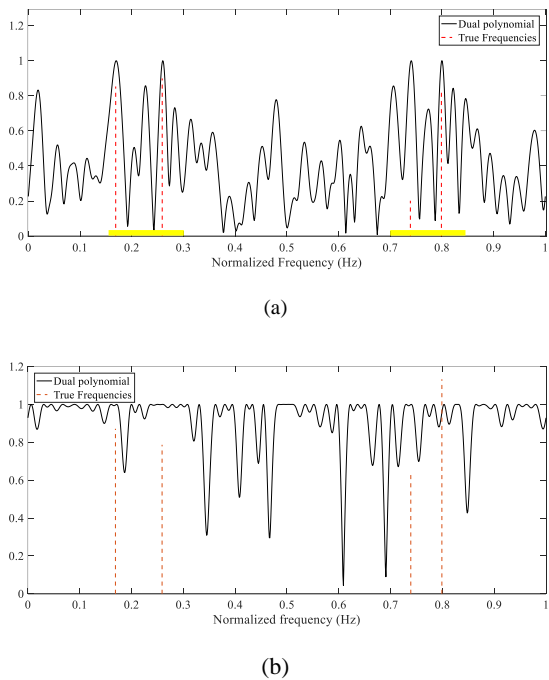

(a)

(b)

Fig.4. Dual polynomial behavior (a) with prior probabilistic ranges and (b) without prior ranges. Prior probabilistic ranges are as [0.15, 0.30] & [0.70, 0.85] (Frequencies are normalized to range [0 1)). Yellow zone defines prior probabilistic range.

We also investigated the behavior of associated dual polynomial of (7), which considers no prior information in reconstruction process. OFDM system and frequencies are the same as before. Fig.4-(b) depicts the results. It is easy to see that presented model can easily identify frequencies while conventional models, even when all of the pilots are used, cannot distinguish true frequencies.

### B. SNR

Second group of simulations were done on the basis of SNR variation from 5 to 30 dB. Because of the popularity of MUSIC algorithm among spectral analysis methods, we compared the presented model with AtomSR and MUSIC. Delay probabilistic priors are the same as before, OFDM system configuration stays the same except that pilot number is fixed at $P=36$. Fig.5. depicts the result of this simulation. It can be seen that with prior knowledge assumption we can greatly reduce the MSE of the estimated channel.

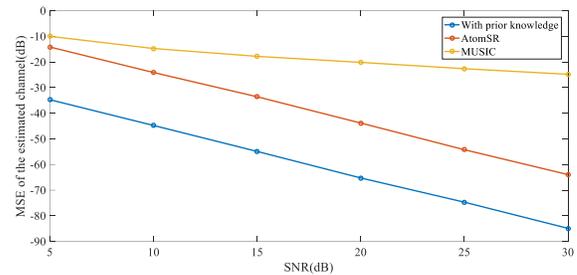

Fig. 5. Comparison of the presented model with AtomSR method assumptions and MUSIC algorithm with different SNRs. It can be seen that prior knowledge greatly improves the performance of channel estimation for different SNRs. Pilot number is fixed at 36.

### C. Minimum Frequency Separation

Another important issue investigated is minimum frequency separation and the Rate of Successful Recovery (RSR). As discussed above, here frequencies can be treated as delays but for convenience we present the results in the terms of frequencies. We define RSR as (26):

$$RSR = \frac{\text{Number of correct frequncies found}}{\text{Number of overal frequencies found}} \quad (26)$$

As expected, with increasing the distance between frequencies RSR should increase to a limit of one. In these simulations OFDM system configuration is as follow: N=256, L=32, pilot allocation scheme is the same as before and pilot number is $P=20$. No noise was considered but results are approximately the same for the noisy case. One prior probabilistic range [0.15, 0.30] and two frequencies were considered. Note that this is equal to the presence of two distinct paths. One frequency was fixed at 0.17 and the other was slowly taking distance. Fig.6. shows the result of this simulation.

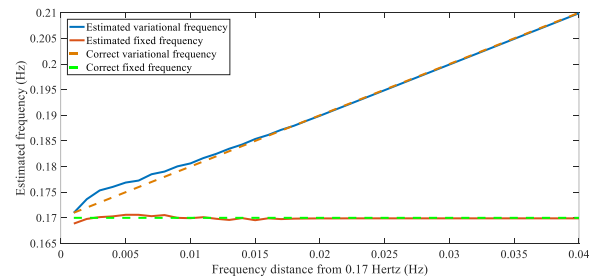

Fig.6. Estimated frequencies when one frequency is fixed and the other is taking distance. Green dashed line presents the correct fixed frequency and the orange dashed line shows correct variational frequency.

As it can be seen from Fig.6, estimated frequencies are meaningless and cannot help detecting the correct frequencies but, as the distance between two frequencies increases correct results appear. Minimum frequency distance needed for precise estimation is $4/N$, as expected. Note that by increasing OFDM system subcarriers we can reach higher resolution estimation results which leads to better time dispersive sparse channel response estimation.

We also investigated RSR deviations. Fig.7. presents RSR deviations with the same simulation configurations as before.

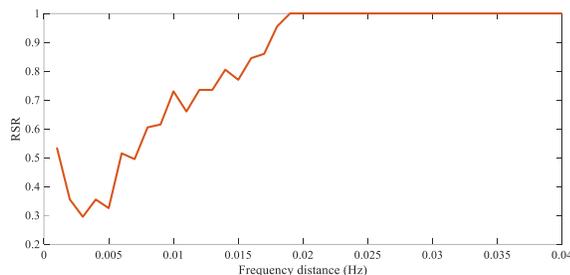

Fig.7. RSR deviations for OFDM system configuration as N=256, L=32, same pilot allocation scheme as before and $P=20$

As Fig.7 illustrates, exact frequency recovery becomes possible only when minimum frequency separation is satisfied. Though superior performance is achieved, more complexity and off course more time are needed to solve the problem (10).

## V. CONCLUSION

Channel estimation in pilot aided OFDM systems with prior knowledge on multipath delays was proposed in this paper. Using weighted atomic norm minimization, priors were imposed on the problem. Simulations showed superior performance compared to AtomSR and MUSIC when using prior knowledge in the estimation problem. It is crystal clear that the benefits of prior knowledge are gained at the cost of complexity and time.